# Theoretical Study on Recognition of Icy Road Surface Condition by Low-THz Frequencies

Xiangzhu Meng, Peian Li, Yuning Hu, Fei Song, and Jianjun Ma

*Abstract*—Recognition of road surface conditions should always be at the forefront of intelligent transportation systems for the enhancement of transportation safety and efficiency. When road surfaces are covered by ice or snow, accident rate would increase due to the reduction of road surface roughness and also friction between tire and road. High-resolution recognition of natural and manmade surfaces has been proved to be achievable by employing radars operating at low-terahertz frequencies. In this work, we present theoretical investigations on surface condition recognition of an icy road by employing low-terahertz frequencies. A theoretical model combining integral equation method (IEM), radiative transfer equation (RTE) and Rayleigh scattering theory is developed. Good agreement between the calculation results and measured data confirms the applicability of low-terahertz frequencies for the evaluation of icy road surface in winter. The influence of carrier frequency, ambient temperature, impurities inside the ice layer and frozen soil surface conditions on the efficiency of this method is presented and discussed.

## I. Introduction

Road surface condition could always be one of the factors that could greatly influence driving safety and can be inferred by the surface parameters, such as roughness, texture, etc. [1]. To achieve high-resolution recognition of road surface conditions, a potential method is employing automotive radars operating at low-terahertz (low-THz) frequencies with the advantages of inherently large bandwidth and short wavelength [2, 3]. Compared with present mm-wave radars, low-THz radars are more sensitive to surface roughness and texture with reduced system mass and packaging issues [4]. Examples of radar images measured using low-THz frequencies could be found in [5-7]. Beside this, low-THz radars are less susceptible to most types of contamination, obscurants and adverse weather particles in outdoor compared to optics [8]. This makes them to match the requirement for future sensing of road surfaces when optical sensing fail in adverse weather conditions, such as snow, rain, water fog, dust fog and so on [9-13].

The analysis of automotive radar's response on road surfaces in adverse weathers require significant experimental and theoretical efforts. The results in [14] and [15] show high recognition of ice/water layer thickness by low-THz frequencies. Reference [16] and [17] investigated the possibility of snow surface monitoring and concluded that low-THz radars can only be used for water content monitoring. Reference [18] showed that the low-THz radars could be used for the monitoring of soil properties, such as mineral fraction, soil particle size and moisture content. However, there is still lack of contributions on practical roads in adverse weathers, such as icy/snowy roads in winter, due to the difficulties in outdoor testing and in emulation of road conditions in laboratories.

To avoid these challenges in experimental measurements, theoretical analysis becomes more important for the improvement of our knowledge on road surface behaviors. In this work, we present investigations on surface recognition of an icy soil road by low-THz frequencies. Similar work has never been conducted before, but we think such a comprehensive study is really necessary for the development of automotive driving technique and the method could be extended to other road types such as asphalt or concrete roads.

The rest of this manuscript is organized as follows. Section II presents the dielectric properties of ice and channel models for the propagation of low-THz frequencies through an ice layer with impurities included. Section III shows dielectric properties of soil and channel models for the low-THz frequencies reflected by soil surfaces at different ambient temperatures. Section IV presents channel models for soil roads covered by ice and investigates the influence of carrier frequency, ambient temperature and soil surface properties. Section V shows the conclusion and implications of this work.

## II. Dielectric properties of ice and channel modeling

The dielectric properties of ice are necessary for the understanding of sensing signatures and there have been many publications concentrated on this topic. Matzler *et al.* [19] performed precise measurements on dielectric loss factors of ice at mainly -15 °C and -5 °C. They concluded that the previous publications showed higher losses and attributed it to the disturbance by impurities inside their samples, such as air bubbles. This is further confirmed by the density discrepancy between pure ice crystal and ice layers. The density of latter one was measured to be 0.909 g /cm$^3$ [20], 0.63-0.95 g/cm$^3$ [21] and

Xiangzhu Meng is studying in Beijing Institute of Technology, Beijing, China.
Peian Li is studying in Beijing Institute of Technology, Beijing, China (e-mail: 3220205091@bit.edu.cn).

Yuning Hu is studying in Beijing Institute of Technology, Beijing, China.
Fei Song is with Beijing Jiaotong University, Beijing, China.
Jianjun Ma, the corresponding author, is with Beijing Institute of Technology, Beijing, China (e-mail: jianjun_ma@bit.edu.cn).

0.4-0.8 g/cm³ [22] due to the existence of air bubbles inside Based on this, Koh [23] fabricated bubble-free ice samples in his lab and observed a lower loss factor than Matzler's measurement, but impurities still exist and disturb the phase relationship of reflected components by the top and bottom sample surfaces. Such impurities are usually considered to be dust and ionic particles. The dust particles do not interfere the dielectric loss factors of ice and snow [24], while the ionic impurities affect that much by leading to the occurring of liquid phase [25], which was observed for snow layers [26].

By reviewing present publications, the Debye dielectric model [27], expressed as $\varepsilon_i = \varepsilon_i' - j\varepsilon_i''$ with $\varepsilon_i' = \varepsilon_{i\infty} + (\varepsilon_{i0} - \varepsilon_{i\infty})/[1 + (2\pi f\tau_i)^2]$ and $\varepsilon_i'' = 2\pi f\tau_i(\varepsilon_{i0} - \varepsilon_{i\infty})/[1 + (2\pi f\tau_i)^2]$, is considered to be the most accurate model for the estimation of dielectric properties of ice. Here, the quantity in the relation of $2\pi f\tau_i = f/f_{i0} \geq 1$ is considered because the relaxation frequency $f_{i0}$ of pure ice occurs in the kilohertz region, which is far below the THz frequency. Consequently, the real and imaginary parts of the permittivity can be simplified to be $\varepsilon_i' \approx \varepsilon_{i\infty}$ and $\varepsilon_i'' = (\varepsilon_{i0} - \varepsilon_{i\infty})/2\pi f\tau_i = \alpha_0/f$ with $\alpha_0 = (\varepsilon_{i0} - \varepsilon_{i\infty})/2\pi\tau_i$. $\varepsilon_i'$ is essentially independent of frequency and exhibits a weak temperature dependence [19] in the form of

$$\varepsilon_i' = 3.1884 + 9.1 \times 10^{-4}T \quad (1)$$

with $T$ being the temperature in °C. Practically, the temperature dependence could be ignored and the value of $\varepsilon_i'$ can be simplified to be a constant as $\varepsilon_i' = 3.1884$. The $\varepsilon_i''$ value varies as the $1/f$ and the coefficient $\alpha_0$ only depends on temperature $T$ as discussed in [28, 29]. However, the general behavior of dielectric loss of ice is characterized by the high frequency tail of the Debye relaxation spectrum with a relaxation frequency in the kilohertz range and by the low frequency tail of the far-infrared absorption bands due to lattice vibrations. So, there is an infrared absorption spectrum that includes a non-resonant term that varies with respect to frequency $f$ [30, 31]. This makes

$$\varepsilon_i'' = \alpha_0/f + \beta_0 f + \gamma_0 f^3 \quad (2)$$

And the superposition of these tails leads to a deep minimum of dielectric loss centered at 2-4 GHz with the values in the order of $10^{-4}$. As explained in [32], the coefficients $\alpha_0$, $\beta_0$ and $\gamma_0$ could be obtained by semiempirical expressions as

$$\alpha_0 = (0.00504 + 0.0062\theta)exp(-22.1\theta) \quad (3\text{-}1)$$
$$\beta_0 = \frac{B_1}{T_K}\frac{exp(b/T_K)}{[exp(b/T_K)-1]^2} + B_2 f^2$$
$$+ exp[-9.963 + 0.0372(T_K - 273.16)] \quad (3\text{-}2)$$
$$\gamma_0 = 1.16 \times 10^{-11} \quad (3\text{-}3)$$

with $\theta = 300/T_K - 1$, $B_1 = 0.0207$ and $B_2 = 1.16 \times 10^{-11}$.

Norouzian *et al.* [15] measured the transmission of low-THz signals through an ice layer. When the ice thickness was increased from 2 cm to 6 cm, the transmissivity would be reduced from -2 dB to -8 dB due to the losses caused by reflections. Koh [23] conducted similar measurements by employing a radar operating at 75-110 GHz and a linear relationship between the carrier frequency and the mean value of measured transmittance was observed. Based on this trend, we could deduce that the transmissivity changes from -1.5 dB to -3.5 dB corresponding to identical ice thickness variations in [15]. These values are much smaller than the that measured by Norouzian, because smooth bubble-free ice slabs were used with the concentration of impurities reduced significantly. We attribute this discrepancy to the volume scattering loss caused by impurities inside the ice layer. Thus, the Fresnel equation and transmission theory employing dielectric properties of pure ice would not be an accurate method, even for the work of Koh, because there are still impurity inclusions inside which leads to the depression of interference patterns.

In this work, to avoid the uncertainties in dielectric values of ice caused by impurities, we would assume a pure ice layer with randomly distributed inhomogeneities located inside. In the model, we just employ the dielectric values of pure ice as in (1) and (2), and introduce the influence of impurities by using a volume scattering theory, where the volume scattering effect could destroy the coherent components and also the phase relationships of them would no longer be preserved. This could be conducted by the radiative transfer equation (RTE) [33] with iterative solution method [34] employed to obtain expressions for downward incidence as in Fig. 1(a), downward emission in Fig. 1(b), upward (c) and downward (d) scattering components.

For the three-layer medium as shown in Fig. 1(a), the top and bottom interfaces are flat with specular reflections considered. In the middle layer (layer 2), uniform distribution of impurities is assumed. The total transmissivity contributed by the incoherent components could be obtained by adding all the components as in Fig. 1 and can be expressed in the form of

$$T = \left(\frac{1-\Gamma_{23}^p}{1-\Gamma_{12}^p\Gamma_{23}^p\gamma^2}\right)\left[(1+\Gamma_{12}^p\gamma)(1-a)(1-\gamma) + (1-\Gamma_{12}^p)\gamma\right] \quad (4)$$

with $a$ representing the single scattering albedo of the ice layer. It could be expressed to be $a = \kappa_s/\kappa_e$ with $\kappa_s$ as the scattering coefficient and $\kappa_e$ as the extinction coefficient evaluated by Mie or Reyleigh scattering theory depending on the comparison of the inclusion size and the THz wavelength [35]. The parameter $\gamma$ is the one-way transmissivity in the pure ice layer and can be express as $\gamma = exp(-d\ sec\theta_2)$ with $d$ being thickness of the ice layer and $\theta_2$ as the refraction angle as shown in Fig. 1(a) corresponding to an incidence angle $\theta_1$. $\Gamma_{12}^p(\theta_1)$ and $\Gamma_{23}^p(\theta_2)$ are the coherent $p$- (V- or H-) polarized reflectivity of the air-ice (1-2) and ice-air (2-1) boundaries at incidence and refraction angles of $\theta_1$ and $\theta_2$, respectively. Both parameters could be obtained by the Fresnel equations [36].

In order to check the efficiency of this model, we estimate the transmissivity of a 150 GHz and a 300 GHz signal through an ice layer with its thickness changes from 0 cm to 7 cm. We regard the impurities to be air bubbles with the value of refractive index being unity. In Fig. 2(a) and (b), the variation of signal transmission versus ice thickness is plotted. The transmissivity decreases seriously for thicker ice layers due to more power losses by absorption and scattering. The calculation agrees well with the measured data obtained by Norouzian *et al.* [15] with normal incidence, who employed a Frequency Modulated Continuous Wave (FMCW) radar to record the wave propagation at normal incidence through a relatively uniform ice layer positioned in an environmental chamber. The agreement indicates that the RTE model is more effective for the analysis of transmissivity than the Fresnel equations with uncertain dielectric values of impure ice considered. This model is very useful for us to improve the knowledge on pure/impure ice behavior on low-THz signal propagation. Fig. 2(c) shows the influence of incidence angle $\theta_1$ on the transmission of THz waves and the transmissivity could be increased by reducing the incidence angle. We attribute this to the smaller reflection coefficient on the medium interfaces (air-ice and ice-air) and the shorter propagation distance inside the ice layer.

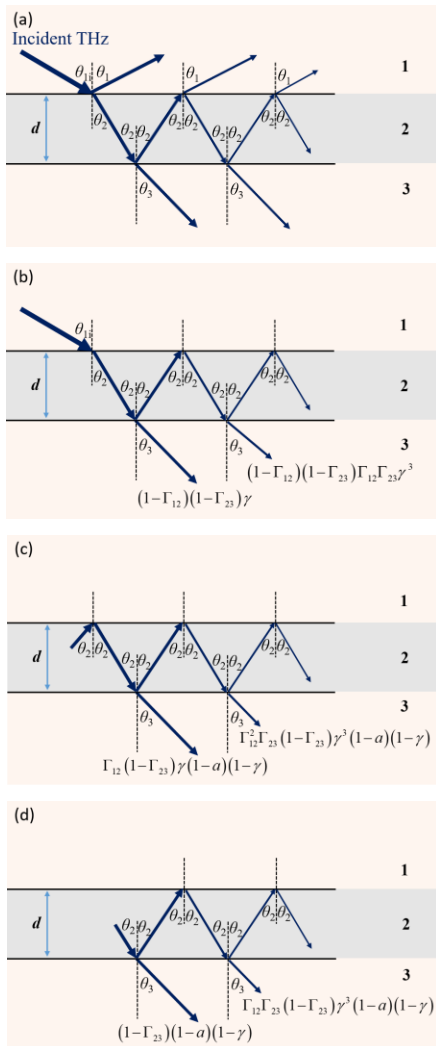

Fig. 1. (a) Basic geometry of transmission through an ice layer. Layer 1, 2, 3 represent air, ice and air layers, respectively. Diagram illustrating of the components of (b) downward propagation from layer 1, (c) upward propagation from layer 2 and (d) downward propagation from layer 2.

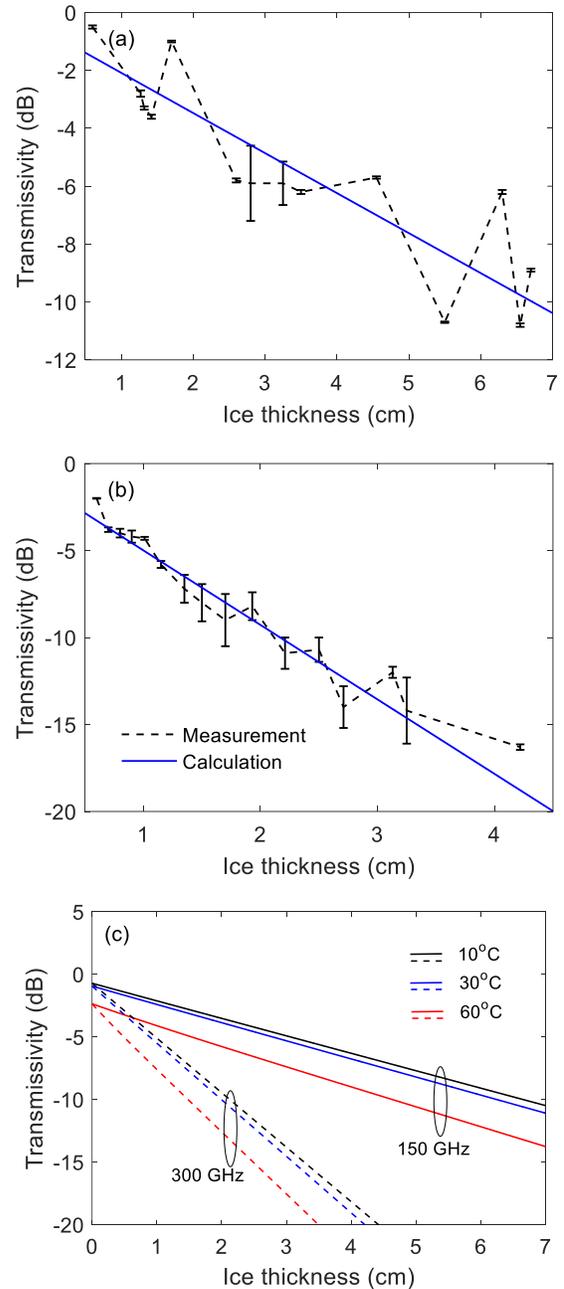

Fig. 2. Comparison of calculated transmissivity versus ice thickness at (a) 150 GHz and (b) 300 GHz to measured data in [15] with the ambient temperature at -7 °C and incidence angle $\theta_1 = 0°$; (c) Variation of calculated hh-polarized transmissivity versus ice thickness at difference incidence angles $\theta_1$.

### III. DIELECTRIC PROPERTIES OF FROZEN SOIL AND CHANNEL MODELING

There have been numerous studies concentrated on the determination of soil dielectric behaviors [37-40]. Generally, a soil medium is electromagnetically a four-component dielectric mixture consisting of bulk soil, bound water, free water and air. The bulk soil is a combination of clay, silt and sand [41], which could be used for soil classification by examining their relative mass fractions [42]. This means the dielectric properties of bulk soil would change with the variation of relative mass fractions of each component as $\varepsilon'_{soil} = (1 + 0.44\rho_b)^2$. Here, the parameter $\rho_b$ is the bulk density of soil, which depends on soil constituents as in Table 1. Bound water represents water molecules very close to soil particles, where the large matric and osmotic forces would make the water molecules to be held tightly [27]. However, when the water molecules are located several molecular layers away from soil particles, the matric forces would become negligible. Thus, the water molecules can move within the soil medium easily and referred to as free water. Both the bound water and free water are commonly used to characterize the moisture content of a soil sample.

According to the Birchak mixing model [43], which is usually used for the evaluation of soil dielectric behavior on THz frequencies, the dielectric constant of soil can be expressed as $\varepsilon_{soil}^{\alpha} = v_s \varepsilon_s^{\alpha} + v_{fw} \varepsilon_{fw}^{\alpha} + v_{bw} \varepsilon_{bw}^{\alpha} + v_a \varepsilon_a^{\alpha}$ [44] based upon refractive volumetric mixing of four components. Here, $\alpha$ is used for model classification. $\alpha = 1$ corresponds to the linear model, $\alpha = 1/2$ to the refractive model, and $\alpha = 1/3$ to the cubic model. The subscripts, $f_w$, $b_w$, $s$, $a$ represent soil, free water, bound water and air, respectively. The terms for free water and bound water can be combined together to get a semiempirical model [45] as

$$\varepsilon_{soil}^{\alpha} = 1 + (\varepsilon_s^{\alpha} - 1) \cdot \rho_b/\rho_s + m_v^{\beta} \varepsilon_{fw}^{\alpha} - m_v \quad (5)$$

to avoid the uncertainty of dielectric constant of bound water. The new parameter $m_v = V_w/V_t \times 100\%$ is the volumetric moisture with $V_w$ as the water volume and $V_t$ as the total volume of the sample, which includes the volumes of air, soil, and water. It is usually preferred than the gravimetric moisture $m_g$ because of its higher sensitivity to water content [46]. Then, the real and imaginary dielectric parts of (5) can be conducted as

$$\varepsilon_{soil}' = [1 + 0.66\rho_b + m_v^{\beta_1}(\varepsilon_w')^{\alpha}]^{1/\alpha} \quad (6\text{-}1)$$
$$\varepsilon_{soil}'' = m_v^{\beta_2} \varepsilon_w'' \quad (6\text{-}2)$$

with $\rho_b$ standing for the bulk density of the soil. The parameter $\varepsilon_w$ is the dielectric constant of water given by Debye model [47] and the values of other parameters can be obtained as

$$\alpha = 0.65 \quad (7\text{-}1)$$
$$\beta_1 = 1.27 - 0.519S - 0.152C \quad (7\text{-}2)$$
$$\beta_2 = 2.06 - 0.928S - 0.225C \quad (7\text{-}3)$$

with $S$ and $C$ being the mass fractions of sand and clay, respectively.

Fig 3(a) shows the contributions by the layers 1, 2 and 3, which represents air, dry soil and the container, respectively. Component (A) is generated by the direct backscattering from top soil surface, (B) by the bottom soil surface and (C) by volume scattering by soil particles. Components (D), (E) are from the volume-surface and surface-volume scattering, respectively. (F) is due to the surface-volume-surface scattering. For the prediction of low-THz frequencies reflected by this structure, a model combining the integral equation method (IEM) [48] and the RTE model is proposed here. The IEM model could be used to estimate the components scattered by rough surfaces (such as metallic rough surface [49], sandpaper surface [50], etc.) by combining the Kirchhoff field and the complementary field as the surface field. The RTE model is responsible for the estimation of the volume scattering effect by the inhomogeneous soil particles. This model has been used for the prediction of THz waves backscattered by dielectric mediums successfully [51]. In this work, we would use it to predict the backscattering performance of low-THz signals by a dry soil layer with a thickness of 4.5 cm. In (4), the value of the single scattering albedo $a$ of the soil layer is obtained by the Rayleigh scattering theory because of the average size of soil particles (80 um) less than about one-tenth the wavelength of the incident low-THz frequencies [35]. The comparison between the measured data in [52] at room temperature (23 °C) and the calculation is shown in Fig. 3(b) and (c). A discrepancy is observed and should be attributed to the wrong characterization of surface roughness according to the demonstration in [53].

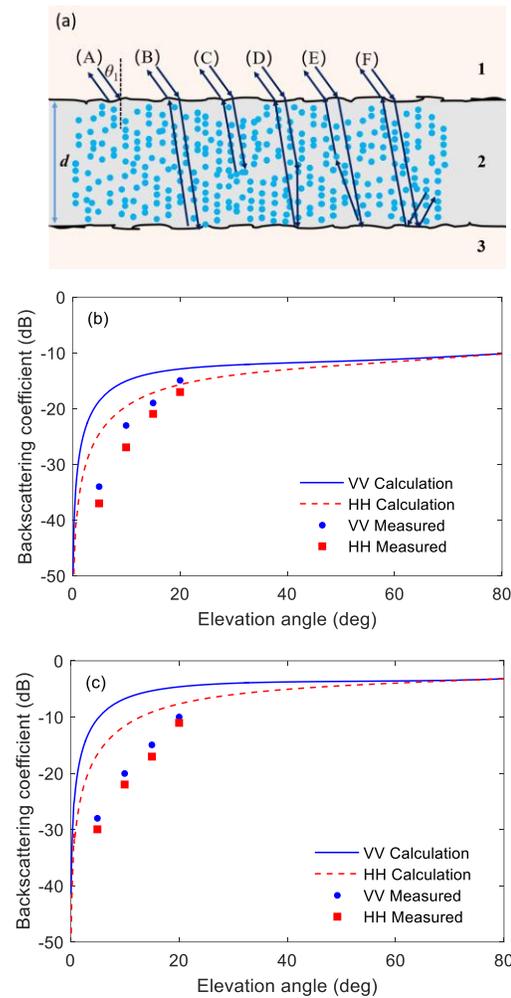

Fig. 3. (a) Scattering contributions by dry soil layer inside a container [34]. Layer 1, 2, 3 represent air, dry soil and the container, respectively. Comparison of calculated backscattering coefficient versus elevation angle (90° - incidence angle) at (b) 100 GHz and (c) 240 GHz to measured data obtain in [52]. (Soil surface RMS height 190 um, surface correlation length 500 um).

For a soil road covered by ice, the soil becomes frozen soil and the above model for unfrozen soil should be modified by introducing the dielectric values of frozen soil. There are seldom publications concentrated on this in THz range. Zhang et al. [41, 54, 55] did some work by inserting two ice-related terms to (5) as

$$\varepsilon_{soil}^{\alpha} = 1 + (\varepsilon_s^{\alpha} - 1) \cdot \frac{\rho_b}{\rho_s} + m_v^{\beta}\varepsilon_{fw}^{\alpha} - m_v + m_{vi}^{\beta}\varepsilon_i^{\alpha} - m_{vi} \quad (8)$$

Where the parameter $m_{vi}$ is the volumetric ice content and can be obtained by expression $m_{vi} = (m_v - m_{vw})\rho_b/\rho_i$ with $\rho_i$ as the specific density of ice in g/cm³. $m_v$ is the total moisture content as we demonstrated early. The term $m_{vw}$ is the unfrozen volumetric moisture content and can be expressed as $m_{vw} = A|T - 273.2|^{-B} \cdot \rho_b/\rho_w$ with $\rho_w$ as the specific density of water in g/cm3 and T as temperature in Kelvin. A and B are parameters relating to the soil texture as shown in Table 1. We rewrite (8) to be

$$\varepsilon'_{soil} = \left[1 + 0.66\rho_b + m_v^{\beta_1}(\varepsilon'_w)^{\alpha} - m_v + m_{vi}(\varepsilon'_i - 1)^{\alpha}\right]^{1/\alpha} \quad (9\text{-}1)$$

$$\varepsilon''_{soil} = m_v^{\beta_2}\varepsilon''_w + m_{vi}\varepsilon''_i \quad (9\text{-}2)$$

with the values of parameters $\alpha$, $\beta_1$, $\beta_2$ and $\sigma$ identical to the definitions in (7). From this equation, we can say that the dielectric properties of ice depend on ambient temperature because the volumetric content of unfrozen water and ice are temperature dependent [41].

TABLE I
SAMPLE DESCRIPTION AND MASS FRACTIONS OF EACH CONSTITUENTS [41].

| Soil sample | Silt clay(*S1*) | Silt loam (*S2*) | Sandy loam (*S3*) |
|---|---|---|---|
| Specific density (g/cm³) | 2.60 | 2.58 | 2.63 |
| Bulk density (g/cm³) | 1.62 | 1.58 | 1.59 |
| A | 11.33 | 5.28 | 2.70 |
| B | 0.62 | 0.57 | 0.61 |
| Mass fraction (100%) | | | |
| Clay | 47.41 | 19.96 | 9.66 |
| Silt | 45.76 | 51.46 | 39.61 |
| Sand | 6.83 | 28.58 | 50.73 |

Particle size of clay is 0.5um to 20 um with a mean size of 6.5 um, silt is 4um to 100um with a mean size of 27 um, sand is 80um to 240 um with a mean value of 140 um [18].

In Fig. 4(a), the real part of permittivity $\varepsilon'$ of frozen soil at 140 GHz is plotted with respect to temperature variation. With the increasing of the temperature, the value of $\varepsilon'$ almost follows a horizontal line firstly because the $\varepsilon'$ values of water and ice are always regarded to be in weak temperature dependence. When the temperature reaches to -2°C, obvious dielectric behavior starts. This is due to the conversion of ice to unfrozen water [56] and the $\varepsilon'$ value of water in the range of 5-5.8 is larger than that of ice around 3.18 as in Fig. 4(a). In Fig. 4(b), the imaginary part of permittivity $\varepsilon''$ of frozen soil (black lines), a temperature dependent value, always changes with the increasing of temperature and a sharp change comes when the temperature closes to the freezing point where the conversion of ice (with the $\varepsilon''$ value of around 0.01) to unfrozen water (with the $\varepsilon''$ value of 3.93-6.36) starts. It should be noted that the soil sample *S1* owns a higher permittivity values in the real and imaginary parts due to a higher fraction of clay it contains as shown in Table 1. As demonstrated in [41], clay is more capable for water storage than sand and silt. Thus, we can say that the sample *S1* would lead to a higher reflection and absorption to the low-THz frequencies.

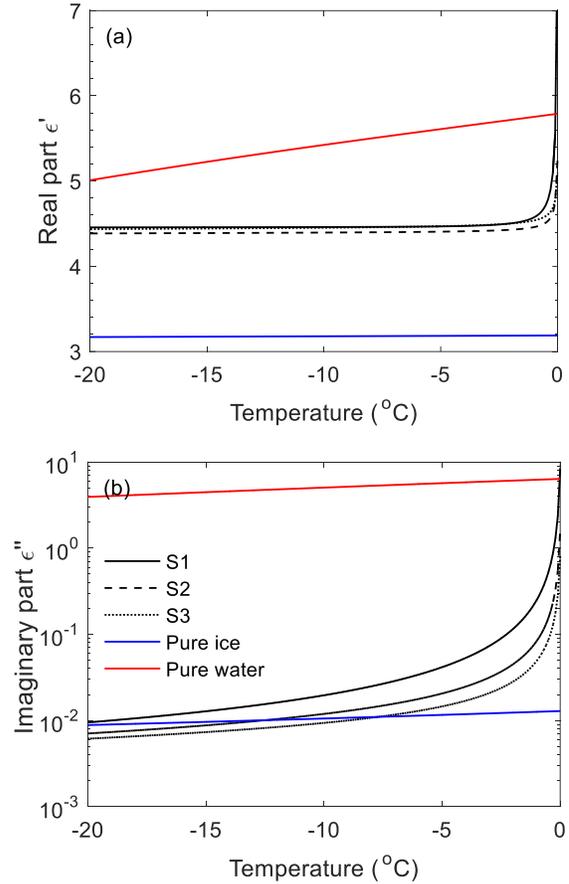

Fig. 4. Variation of dielectric properties of frozen soil, pure ice and water to 140 GHz with respect to temperature. (Total water content $m_v = 30\%$, *S1*: silt clay, *S2*: silt loam, *S3*: sandy loam).

By employing the IEM and RTE theories again and using the parameters of silt clay (Sample *S1*) in Table 1, the backscattering coefficient by frozen soil with a surface RMS height of 190 um and surface correlation length of 500 um is calculated and shown in Fig. 5. In Fig. 5(a), the evolution of backscattering coefficient versus elevation angle under different temperatures is presented when the radar operates at 140 GHz. It is obvious that, the backscattering coefficient decreases with the increasing of temperature over the whole elevation angle range from 0° to 90°, due to the temperature dependence of the loss factor (the $\varepsilon''$ value). And the influence of temperature is reduced when the radar signal is launched at a large elevation angle (corresponding to a smaller incidence angle $\theta_1$). Thus, a small incidence angle should be set in measurements to minimize the influence of temperature. In Fig. 5(b), we compare the backscattering coefficients by all the soil samples in Table 1 and find that sample *S1* presents the minimum backscattering due to its higher loss factor as shown in Fig. 4(b). It should also be noted that, higher scattering coefficient could be achieved by the employing the VV polarization component.

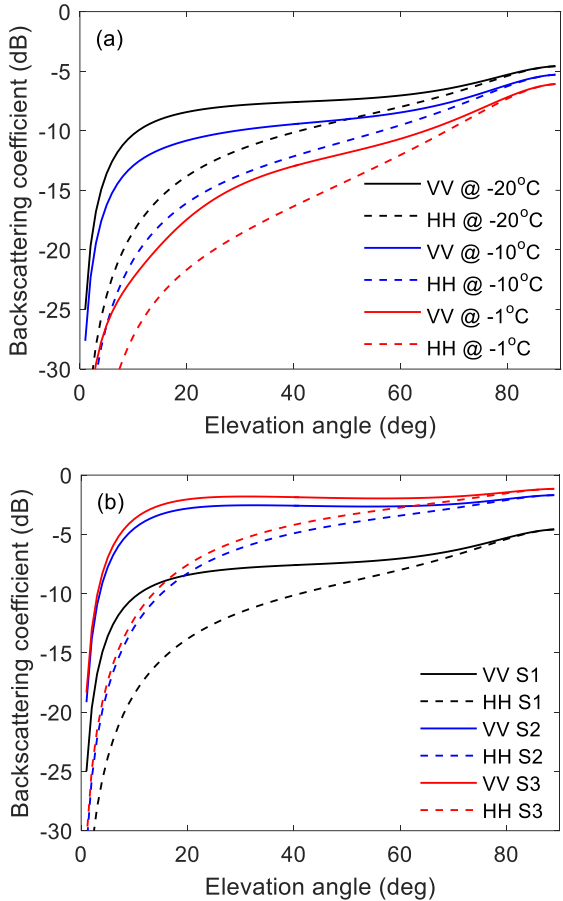

Fig. 5. (a) Evolution of backscattering coefficient of frozen Sample *S1* (silt clay) to 140 GHz versus elevation angle (90° - incidence angle) under different temperatures; (b) Comparison of backscattering behaviors by different frozen samples on a 140 GHz signal at *T* = -20°C. (Frozen soil surface RMS height $S_3$ = 190 um, surface correlation length $L_3$ = 500 um, total moisture content $m_v$ = 30%)

## IV. CHANNEL MODELING ON ICY SOIL ROAD

To evaluate the performance of low-THz frequencies caused by an icy soil road, we combine theoretical models for the ice layer and frozen soil layer together. Then the layer 3 becomes a frozen soil layer. Naturally, there should be more air bubbles inside the ice layer, which would lead to its density being smaller than pure ice, such as a value of 0.909 g/cm³ measured in [20]. Calculations are conducted by using parameters in Table II and results are plotted in Fig. 6. Fig. 6(a) shows a positive correlation between the backscattering coefficient and thickness of the ice layer above frozen soil. When the ice thickness increases, the influence of volume scattering grows and the backscattering coefficient increases. This means low-THz frequencies could be capable for recognition of ice layer thickness. It should also be noted that the thickness range recognized by the 140 GHz signal is 0 cm to 4 cm, while that becomes 0 cm to 1 cm for the 220 GHz signal and 0 mm to 0.3 mm for the 340 GHz signal. Comparing the slopes of the curves in these ranges, the 340 GHz signal owns the largest slope, which means the resolution could be by employing higher carrier frequencies. Thus, we can say that 140 GHz frequency could offer a wider thickness recognition range, while the 340 GHz frequency owns a higher resolution.

Fig. 6(b) shows the evolution of backscattering coefficient of a 140 GHz signal with respect to ice surface (corresponding to the 1-2 interface in Fig. 1) conditions represented by surface RMS height $S_1$ and correlation length $L_1$. The surface RMS height is a standard deviation of the surface height, which stands for the roughness of a surface. The correlation length represents the correlation between two adjacent points of a surface and indicates surface homogeneity. Both parameters are attributed to frictions since they define the height and correlation of the ice asperities which interact with auto tire [57]. Ref. [58] shows the variation of friction coefficient between ice and tire with respect to surface roughness. At contact pressure 0.30 MPa, temperature -10°C and ice surface roughness 0.11 mm, the friction coefficient between both is around 0.3, which leads the accident risk 300 times higher than that when the friction coefficient is 0.6 [59]. So in our calculation, we set the RMS roughness ranging from 0.01 mm to 0.5mm.

The positive correlation relationship between the scattering coefficient and the RMS height $S_1$ indicates that the 140 GHz signal is efficient to monitor the surface roughness of the ice layer. When we increase the ice surface correlation length $L_1$, the slope of the curves is reduced, which means the resolution is reduced also.

TABLE II
LIST OF PARAMETERS EMPLOYED IN CALCULATION

| Parameter | Value |
|---|---|
| Carrier frequency *f* | 140 GHz |
| Incidence angle $\theta_1$ | 10 ° |
| Ambient temperature *T* | -10 °C |
| Pressure *P* | 1013 hPa |
| Thickness of ice layer *d* | 0.5 cm |
| Frozen soil type | Silt clay (*S1*) |
| Total water content $m_v$ | 30% |
| Ice surface RMS height $S_1$ | 0.2 mm |
| Ice surface correlation length $L_1$ | 50 mm |
| Frozen soil surface RMS height $S_3$ | 0.19 mm |
| Frozen soil surface correlation length $L_3$ | 0.5 mm |

Dependence of the backscattering behavior of the 140 GHz signal on the temperature variation is shown in Fig. 6(c). In the temperature range of -30 °C to -2 °C, the horizontal lines indicate the negligible influence of ambient temperature. We attribute this to the weak temperature dependence of dielectric permittivity of ice (as in Fig. 4) which affects the backscattering behavior more than that of frozen soil. However, the friction coefficient between ice and tire would decrease linearly with the increasing of temperature due to the positive temperature dependence of yield stress, which is defined as the minimum stress required to cause flow [60] and is still not fully understood even though many experimental research works have been conducted [58]. When the temperature increases to -2 °C and above, obvious change comes due to the conversion of ice to unfrozen water. Compare the amplitude of the horizontal lines under different values of ice thickness $d$ and ice surface RMS height $S_1$. It could be concluded again that the method on recognition of ice thickness and roughness by low-THz frequencies is efficient.

To see the influence of frozen soil surface conditions, such as surface RMS height $S_3$ and correlation length $L_3$, we plot the evolution of backscattering coefficient of a 140 GHz frequency and show it in Fig. 6(d). It can be seen that both parameters would affect the backscattering behavior of the signal obviously. Higher correlation length $L_3$ leads to higher backscattering effect, which enforces the recognition efficiency of the icy road surface.

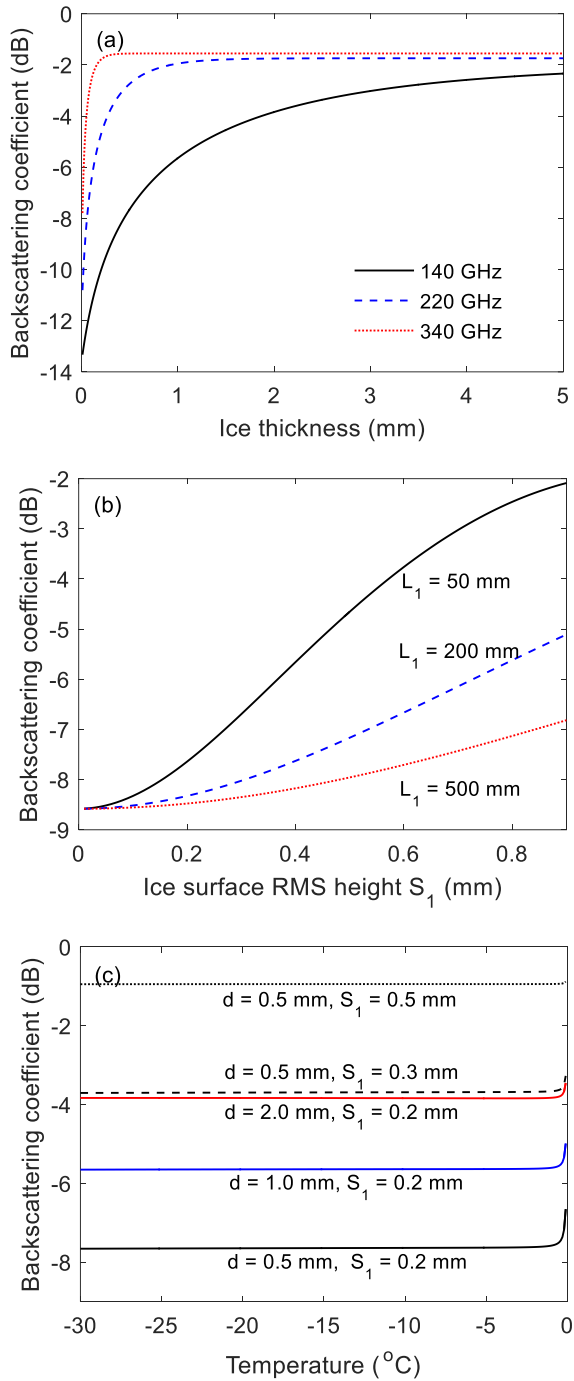

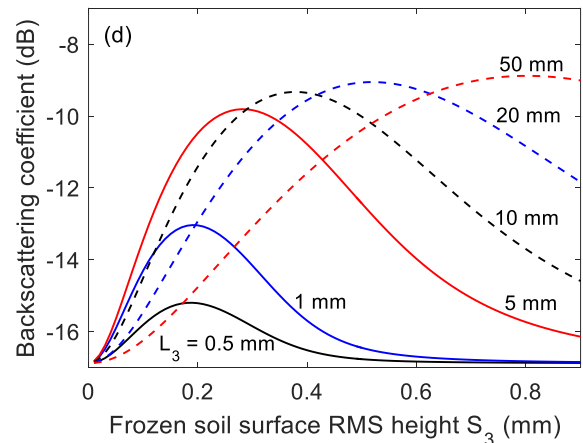

Fig. 6. Variation of backscattering coefficient (a) versus ice thickness for the THz wave at 140 GHz, 220 GHz and 340 GHz; (b) versus ice surface RMS height $S_1$ under different correlation length $L_1$ values; (c) versus temperature under different ice thickness $d$ values and ice surface RMS height $S_1$ values; (d) versus frozen soil surface RMS height $S_3$ under different correlation length $L_3$ values. (Other parameters employed are shown in Table II with incidence angle $\theta_1 = 10$ºC).

## V. CONCLUSIONS

In this paper, we investigate the capability of low-THz frequencies for the recognition of icy road surfaces due to their high sensitivity to slight surface roughness. A theoretical model, including surface and volume scattering mechanisms, is proposed to evaluate the backscattering behavior and is examined by employing measured data. The influence of impurities inside the ice layer, ambient temperature, carrier frequency and frozen soil surface conditions on the backscattering behavior of the signals is investigated. Low-THz frequencies are identified to be capable for the recognition of ice layer thickness and surface roughness of a soil road, even though the efficiency would be affected by the carrier frequency and the soil road surface conditions. This work confirms the current and near-future applications of low-THz frequencies in road surface recognition and the developed theoretical model could also be suitable for the surface evaluation of icy/snowy concrete road or asphalt road.

AUTHORSHIP CONTRIBUTION STATEMENT

**Xiangzhu Meng:** Methodology, Calculation, Investigation. **Peian Li:** Methodology, Calculation, Validation, Investigation, Writing. **Ningyu Hu:** Calculation, Validation. **Fei Song:** Investigation, Validation. **Jianjun Ma:** Conceptualization, Methodology, Funding acquisition, Writing, Supervision.

DECLARATION OF COMPETING INTEREST

The authors declare that they have no known competing financial interests or personal relationships that could have appeared to influence the work reported in this paper.

ACKNOWLEDGMENTS

This research was supported by National Natural Science Foundation of China (No. 6207106), Beijing Institute of

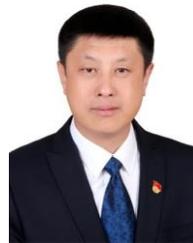

**Xiangzhu Meng** was born in Harbin, China, in 1971. He received the M.S. degree in Electronics and Communication Engineering from Southwest University of Science and Technology, Mianyang, China in 2015. He is now pursuing his Ph.D. degree in Navigation Guidance and Control at Beijing Institute of Technology from 2019.

He is working as the deputy secretary of China's Next Generation Internet (Ipv6) Industry Technology Innovation Strategic Alliance, associate director of Beidou IoT Things Committee of GLAC, council member of Mianyang Entrepreneurship Promotion Association and counselor of Chinese Academy of Physical Engineering

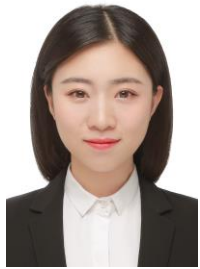
**Peian Li** was born in Tianjin, China, in 1992. She graduated from Minzu University of China, majoring in communication engineering. She graduated from Beijing Jiaotong University with a master's degree in electronic science and technology, and her research area is deep learning. Now she is studying in Beijing Institute of Technology, majoring in electronic information, and the research direction is terahertz time domain detection.

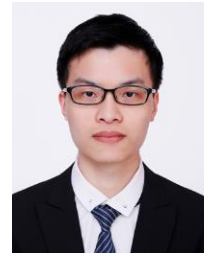
**Yuning Hu** was born in Yiyang, China, in 2000. At present, he is a senior student, majoring in electronic and information engineering at Beijing Institute of Technology, Beijing, China. His current research interest is terahertz channel modeling.

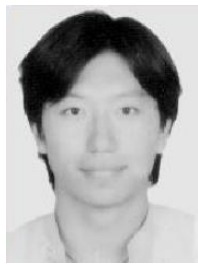
**Fei Song** is a full Professor with the National Engineering Laboratory for Next Generation Internet Technology and the School of Electronic and Information Engineering, Beijing Jiaotong University. His current research interests include network architecture, system security, protocol optimization, and cloud computing. He serves as a technical reviewer for several journals including IEEE Communications Magazine, IEEE Internet of Things Journal, IEEE Trans. on Industrial Informatics, IEEE Trans. on Services Computing, IEEE Trans. on Emerging Topics in Computing, etc.

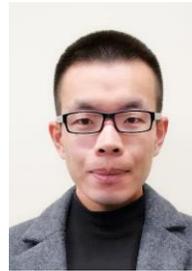
**Jianjun Ma** was born in Qingdao, China, in 1986. He received the Ph.D. degree in applied physics from New Jersey Institute of Technology, Newark, NJ, USA and in 2015, under the guidance of Prof. John F. Federici. His Ph.D. dissertation was about the weather impacts on outdoor terahertz and infrared wireless communication links.

In 2016, he joined Prof. Daniel M. Mittleman's group at Brown University, RI, USA, as a postdoctoral research associate. In 2019, he joined Beijing Institute of Technology, Beijing, China, as a professor. His current research interests include terahertz and infrared wireless communications, terahertz time domain detection, terahertz waveguides, terahertz indoor/outdoor propagation and link security.